\documentclass[a4paper, oneside, 10pt]{article}
\usepackage[T1]{fontenc}
\usepackage[latin1]{inputenc}
\usepackage[english]{babel}
\usepackage[pdftex]{graphicx}
\usepackage{authblk}
\usepackage[round, sort&compress, authoryear]{natbib}
\usepackage{amssymb, amsmath}
\usepackage{bm}
\usepackage{textcomp}
\usepackage{color}
\usepackage{lmodern}

\title{\textsc{Electron Acceleration in Underdense Plasmas Described with a Classical Effective Theory}}
\author[1a]{M.A.~Pocsai}
\author[1b, 2]{S.~Varró}
\author[1c, 2]{I.F.~Barna}
\affil[1]{Wigner Research Centre for Physics of the Hungarian Academy of Sciences\authorcr
			Konkoly--Thege Miklós út 29-33.\\
			H-1121 Budapest, XII.}
\affil[2]{ELI-HU Nonprofit Ltd.\authorcr
			Tisza Lajos krt.~85-87.\\
			H-6728 Szeged}
\affil[a]{pocsai.mihaly@wigner.mta.hu}
\affil[b]{varro.sandor@wigner.mta.hu}
\affil[c]{barna.imre@wigner.mta.hu}
\date{2014 September}

\newcommand{\parenth}[1]{\left( #1 \right)}
\newcommand{\bparenth}[1]{\left[ #1 \right]}

\newcommand{\abs}[1]{\left\lvert #1 \right\rvert}

\newcommand{\diff}{\mathrm{d}}

\begin{document}
\maketitle

\begin{abstract}
An effective theory of laser--plasma based particle acceleration is presented. Here we treated the plasma as a continuous medium with an index of refraction $n_{m}$ in which a single electron propagates. Because of the simplicity of this model, we did not need to perform PIC simulations in order to study the properties of the electron acceleration. We studied the properties of the electron motion due to the Lorentz force and the relativistic equations of motion were numerically solved and analysed. We compared our results to PIC simulations and experimental data.

\end{abstract}

\section{Introduction}
In an early work \cite{Tajima} showed that strong laser fields are capable to create large amplitude waves in plasmas due to the non-linear ponderomotive force. They suggested two methods for electron acceleration in plasmas: the first one is based on beating two monochromatic laser pulses and the second one needs the usage of ultra-intense monochromatic pulses. These methods are called Laser Plasma Beat Wave Acceleration (PBWA) and Laser Wakefield Acceleration (LWFA), respectively. Their studies were based both on analytic calculations and computer simulations. These wakes are responsible for the high energy gain of the accelerated electrons. These basic ideas opens a new horizon of building compact particle accelerators. The key point is the resonant excitation of the plasma in order to create large amplitude plasma waves. In PBWA the frequency of the beat wave, which is the difference of the laser frequencies $\omega_{1}$ and $\omega_{2}$ has to match the plasma frequency: $ \omega_{1} - \omega_{2} = \omega_{p}, $ while in LWFA the laser envelope length $a^{2}$ is on the order of the plasma wavelength $\lambda_{p}.$ In the early '80s only the PBWA method was accessible in the experiments. However in the middle of the '80s a remarkable improvement has been made: \cite{cpa2} invented the chirped pulse amplification (CPA) method which made possible to reach higher and higher laser intensities without damaging the lasing medium. In the early '90s ultra-high intensities $(\geq 10^{18} \, \mathrm{W}/\mathrm{cm^{2}})$ began to be accessible for experiments. Besides, we mention that various schemes have been proposed for generating ultrashort electron pulses (e.g.~\cite{varro-attosecond-2008, lifschitz-proposed-2006}).

The invention of the CPA method has resulted in a considerable development of the plasma based particle accelerators. Numerous reviews and experimental studies have been published in the last two decades. We mention the works of \cite{Esarey2, Malka, Nakajima, Geddes, Gonsalves}. In theoretical plasma physics the popular Particle-in-Cell (PIC) simulation approach became predominant and an essential tool for modelling plasma based particle accelerators. We should also mention the important works by \cite{rosenzweig-acceleration-1991, Vieira, Pukhov}. A robust, versatile state-of-the-art PIC code, called OSIRIS, has been also developed by \cite{OSIRIS}.

In fact, there are alternative ways for modelling electron acceleration in strong electromagnetic fields. One way is the direct integration of the relativistic equations of motion, as it has been done by \cite{Wang1, Wang2}, for instance. For some special cases the equations can be integrated analytically, however, in general the equations can be solved only by numerical means. A very important case is---from both theoretical and experimental point of view---the purely laser based particle acceleration. In this scheme the electrons are accelerated by the strong electromagnetic field of Gaussian laser pulses,  as presented by \cite{Sohbatzadeh1, Sohbatzadeh2}. Even in the frame of relativistic quantum mechanical description \cite{VarroNm} has recently shown that there exist closed analytic solutions of the wave equations (Dirac and Klein-Gordon equations) if one takes into account the effect of the plasma through a phenomenological index of refraction ($n_{m}$). Needless to say, such an approach may also be meaningful in the classical (non-quantum) regime, where the index of refraction is contained in the Lorentz force. We note that the single-particle dynamics in the presence of collective radiation back reaction has also been studied by \cite{varro-linear-2007} for describing high-harmonic generation on plasma layers in the relativistic domain.


In the following study the acceleration mechanisms of a single electron with strong electromagnetic fields embedded in plasmas are investigated. The joint interaction of the electron with the plasma background and with the radiation field is taken into account by a phenomenological index of refraction. So the obtained effective theory is completely based on classical electrodynamics. In Section \ref{sec:theory}, at first we discuss the relativistic equation of motion of the electron in plane electromagnetic radiation. As a second case, a more realistic description is analysed where the spatial extent of the laser pulse has been taken into account. The results are presented and interpreted in Section \ref{sec:results}. Our paper ends with a short summary.

For a better transparency SI units are used throughout the paper if not stated otherwise.

\section{Theory}\label{sec:theory}
In the presence of an electromagnetic field the Lorentz-force acts on the electron:
\begin{equation}
	\mathbf{F} = e\parenth{\mathbf{E} + \mathbf{v} \times \mathbf{B}}
\end{equation}
with $e$ the electron charge, $\mathbf{E}$ the electric field, $\mathbf{B}$ the magnetic field, $\mathbf{v}$ the velocity of the electron and $\mathbf{F}$ the Lorentz-force. The equations of motions for a relativistic electron are the following:
\begin{subequations}\label{eqgrp:Newton--Lorentz}
\begin{align}
	\frac{\diff \mathbf{p}}{\diff t}& = e\parenth{\mathbf{E} + \frac{\mathbf{p}}{m_{e} \gamma} \times \mathbf{B}},\\
	\frac{\diff \gamma}{\diff t}& = \frac{1}{m_{e} c^{2}} \mathbf{F} \cdot \mathbf{v}.
\end{align}
\end{subequations}
It is important to note that one cannot write any arbitrary function in place of $\mathbf{E}$ and $\mathbf{B}$ since the electromagnetic field has to satisfy the electromagnetic wave equation. Hence, the most general form of $\mathbf{E}$ and $\mathbf{B}$ is as follows:
\begin{align}
	\mathbf{E}(t, \mathbf{r})& = \pmb{\varepsilon} E_{0} f \bparenth{\omega \Theta \parenth{t, \mathbf{r}}},\label{eq:EM_general_E}\\
	\mathbf{B}(t, \mathbf{r})& = \mathbf{n} \times \mathbf{E}(t)\label{eq:EM_general_B}
\end{align}
with $\pmb{\varepsilon}$ the polarization vector, $E_{0}$, $\omega$ and $\mathbf{n}$ the amplitude, angular frequency and the unit vector of the propagation of the electromagnetic field and $f$ an arbitrary, smooth function, respectively. For a better transparency, we introduced the following notation:
\begin{equation}\label{eq:Theta}
	\Theta(t, \mathbf{r}) := t - \mathbf{n} \cdot \frac{\mathbf{r}}{c}.
\end{equation}
Note that the components of the electromagnetic field ($\mathbf{E}$ and $\mathbf{B}$) depend only on $\Theta$.

If we also want to take into account the presence of a medium with an index of refraction $n_{m} < 1$, we need to generalize the definition of $\Theta(t, \mathbf{r})$ in the following way:
\begin{equation}\label{eq:Theta-gen}
	\Theta(t, \mathbf{r}, n_{m}) := t - n_{m} \mathbf{n} \cdot \frac{\mathbf{r}}{c}.
\end{equation}
Since the index of refraction of plasmas is less than unity the generalized definition of $\Theta$ naturally describes the situation in which an electron propagates in a plasma. The refraction index of the plasma depends only on the plasma frequency:
\begin{equation}
	n_{m} = \sqrt{1 - \frac{\omega_{p}^{2}}{\omega_{L}^{2}}}
\end{equation}
with
\begin{equation}
	\omega_{p}^{2} = \frac{n_{e}e^{2}}{\varepsilon_{0}m_{e}}
\end{equation}
with $n_{e}$ the electron density in the plasma, $\varepsilon_{0}$ the permittivity of vacuum and $m_{e}$ the electron mass.

Now we shortly summarize the effective theory of electron acceleration in plasmas. Depending on the plasma density we determine $\omega_{p}$ and $n_{m}$. We treat the electromagnetic field, defined with equations \eqref{eq:EM_general_E} and \eqref{eq:EM_general_B} as a function of $\Theta(t, \mathbf{r}, n_{m})$ and solve the relativistic equations of motion \eqref{eqgrp:Newton--Lorentz} numerically. The special case $n_{m} = 1$ describes the electron acceleration with electromagnetic waves---with lasers, for instance. Worthy of note that \cite{VarroExact} showed that the relativistic equations of motion can be integrated exactly with $n_{m} = 1$ for a circularly polarized, linearly chirped electromagnetic plane wave. \cite{Sohbatzadeh1} showed analytic solution for linear polarization. In both cases the solutions can be expressed with the Fresnel integrals; see \cite{Abramowitz}.

In the present work we investigate the acceleration of a single electron with a chirped electromagnetic plane wave with linear polarization, sine-square shaped temporal envelope ($n_{m} = 1$):
\begin{equation}\label{eq:planewave}
	f \bparenth{\Theta(t, \mathbf{r}, n_{m})} = \left\lbrace \begin{array}{ll}
		\sin^{2} \bparenth{\frac{\pi \Theta(t, \mathbf{r}, n_{m})}{T}} \times &\\
		\sin \bparenth{\omega \Theta(t, \mathbf{r}, n_{m}) + \sigma \Theta^{2}(t, \mathbf{r}, n_{m}) + \varphi}& \textrm{if $t \in \bparenth{0, T}$}\\
		0& \textrm{otherwise}
	\end{array}
	\right.
\end{equation}
with $T$ the pulse duration, $\sigma$ the chirp parameter and $\varphi$ the carrier--envelope phase. The electric field itself is polarized in the $x$ direction and propagates in the $y$ direction, that is, $\pmb{\varepsilon} = \mathbf{e}_{x}$, $\mathbf{n} = \mathbf{e}_{y}$. A series of plane wave pulses is given with a sum of single plane waves:
\begin{equation}
	\sum_{l=0}^{N} f \bparenth{\Theta(t, \mathbf{r}, n_{m}) - l T},
\end{equation}
which has to substituted into \eqref{eq:EM_general_E}. As we mentioned, in case of plasma based acceleration $n_{m} < 1$ has to be applied.

For a more realistic description, instead of sine-square envelope we used a Gaussian pulse shape. In this case the formulation for the most general form of the electromagnetic field \eqref{eq:EM_general_E}--\eqref{eq:EM_general_B} cannot be applied conveniently. Hence we give the $x$, $y$ and $z$ components explicitly for an $x$-polarized Gaussian pulse that propagates in the $z$ direction. According to the works by \cite{Sohbatzadeh1, Sohbatzadeh2} the components of such an electric field are:
\begin{subequations}\label{eqgrp:Gauss_E}
\begin{align}
	\begin{split}
	E_{x}& = E_{0} \frac{W_{0}}{W(z)} \exp \bparenth{-\frac{r^{2}}{W^{2}(z)}} \exp \bparenth{-\frac{\Theta^{2}(t, \mathbf{r}, n_{m})}{T^{2}}} \times \\
	& \quad \cos \bparenth{\frac{2kr^{2}}{2 R(z)} - \Phi(z) + \omega \Theta(t, \mathbf{r}, n_{m}) + \sigma \Theta^{2}(t, \mathbf{r}, n_{m}) + \varphi}
	\end{split}\\
	E_{y}& = 0\\
	\begin{split}
	E_{z}& = -\frac{x}{R(z)}E_{x} +\\
	& \quad E_{0} \frac{2x}{k W^{2}(z)} \cdot \frac{W_{0}}{W^{2}(z)} \exp \bparenth{-\frac{r^{2}}{W^{2}(z)}} \exp \bparenth{-\frac{\Theta^{2}(t, \mathbf{r}, n_{m})}{T^{2}}} \times \\
	& \quad \sin \bparenth{\frac{2kr^{2}}{2 R(z)} - \Phi(z) + \omega \Theta(t, \mathbf{r}, n_{m}) + \sigma \Theta^{2}(t, \mathbf{r}, n_{m}) + \varphi}
	\end{split}
\end{align}
\end{subequations}
and the magnetic field is given by
\begin{subequations}\label{eqgrp:Gauss_B}
\begin{align}
	B_{x}& = 0\\
	B_{y}& = E_{x}\\
	\begin{split}
	B_{z}& = \frac{y}{R(z)}E_{x} + E_{0} \frac{2y}{k W^{2}(z)} \cdot \frac{W_{0}}{W^{2}(z)} \exp \bparenth{-\frac{r^{2}}{W^{2}(z)}} \exp \bparenth{-\frac{\Theta^{2}(t)}{T^{2}}} \times \\
	& \quad \sin \bparenth{\frac{2kr^{2}}{2 R(z)} - \Phi(z) + \omega \Theta(t) + \sigma \Theta^{2}(t) + \phi}
	\end{split}
\end{align}
\end{subequations}
with $W_{z} = \bparenth{1 + (z/z_{R})^{2}}^{1/2}$ the beam waist, $R(z) = z \bparenth{1 + (z_{R}/z)^{2}}$ the radius of curvature, $\Phi(z) = \tan^{-1} (z/z_{R})$ the Guoy phase, $W_{0} = \sqrt{\lambda z_{R} / \pi}$ the half of the focused spot size, $z_{R}$ the Rayleigh length, $\varphi$ the carrier--envelope phase, $\lambda$ the wavelength of the laser and $T$ the pulse duration.

\section{Results}\label{sec:results}
First we introduce the results obtained from the electron acceleration in a chirped electromagnetic plane wave. In this case the electromagnetic field is given by equations \eqref{eq:EM_general_E}, \eqref{eq:EM_general_B} and \eqref{eq:planewave}. We investigated the dependence of the electron energy gain on the parameters, i.e.~the initial momentum of the electron, the chirp parameter, the carrier--envelope phase and the field strength of the electromagnetic fields. After that we optimized the parameters in order to get the maximal gain. The numerical solution of the equations \eqref{eqgrp:Newton--Lorentz} have been obtained by using Wolfram \textit{Mathematica}$^{\textrm{\textcopyright}}$ [Copyright 1988--2012 Wolfram Research, Inc.]. The tolerance was set to $10^{-5}$ both for the relative and absolute errors. Complete technical details can be found in the Master's Thesis written by \cite{Pocsai}.

The energy gain is defined as follows:
\begin{equation}
	\Delta E := m_{e} c^{2} \bparenth{\gamma(t = T) - \gamma(t = 0)}.
\end{equation}
which is the difference of the final and the initial kinetic energy. Without chirp we did not experience any energy gain. The energy of the electron varied periodically in time with $T$ period. By positive chirp the energy gain was negligibly small, hence we only took negative values for the chirp parameter. The significant gain via negative chirp can be explained graphically in two different ways: the down-chirp can be interpreted as a down-conversion. Due to the conservation of energy, the decrement of the laser frequency results in an increment in the electron energy. From an other point of view, it can be seen that any kind of chirp causes a significant effect if the chirped pulse contains only a few optical cycles. It can also be seen that the oscillations of the field strength at the front of the pulse are approximately the same of the order of magnitude, but this difference increases in time. At the end of the pulse, the amplitude of the last oscillation is significantly smaller than that of the last but one. That is, the sharp, rising edge in the field strength cannot be compensated. \cite{Cheng} discussed this phenomenon in details.

It is important to note that for acceleration in an electromagnetic plane wave the energy gain of the electron does not depend on the initial position, it depends only on the initial momentum and the parameters. Without loss of generality the choice $p_{z} \equiv 0$ can be made. Because of simplicity we chose the electron to be started from the origin.

\begin{figure}
\begin{center}
	\includegraphics[scale=0.4]{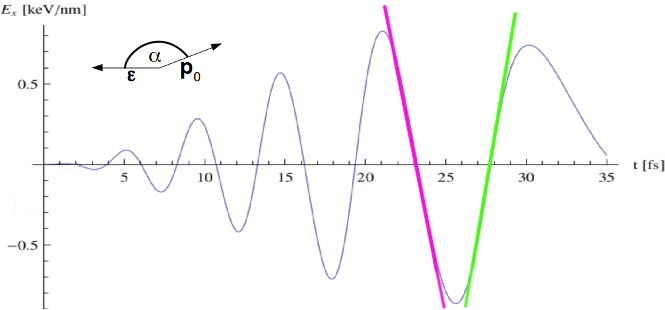}
	\emph{\caption{The $x$-component of the chirped electric field. $\lambda = 800 \, \mathrm{nm}$, $T = 35 \, \mathrm{fs}$, $I = 10^{17} \, \mathrm{W} \mathrm{cm}^{-2}$, $\sigma = -0.03886 \, \mathrm{fs}^{-2}$, $\varphi = 0$. The optimal initial momentum is $\mathbf{p}_{0} = (-1570 \, \mathrm{keV/c}, 450 \, \mathrm{keV/c}, 0)$; $\alpha = 164^{\circ}$ is the ideal injection angle, related to the positive $x$ direction. Note the sharp falling and rising edges (highlighted in purple and green, respectively): they are mainly responsible for the acceleration.}\label{fig:chirped}}
\end{center}
\end{figure}

\begin{figure}
\begin{center}
	\includegraphics[scale=0.6]{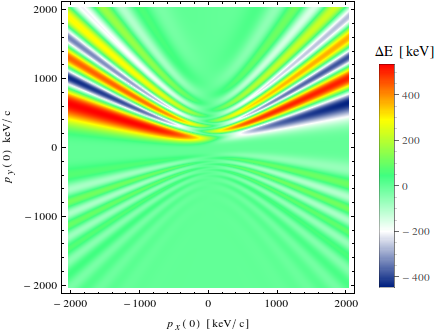}
	\emph{\caption{The dependence of the energy gain on the initial momentum. $\lambda = 800 \, \mathrm{nm}$, $T = 35 \, \mathrm{fs}$, $I = 10^{17} \, \mathrm{W} \mathrm{cm}^{-2}$, $\sigma = -0.03886 \, \mathrm{fs}^{-2}$, $\varphi = 0$. The optimal initial momentum is $\mathbf{p}_{0} = (-1570 \, \mathrm{keV/c}, 450 \, \mathrm{keV/c}, 0)$. Note that the maximal energy gain is obtained if the initial direction of the electron is $164 ^{\circ}$ relative to the positive $x$ direction (see also Fig.~\ref{fig:chirped}).}\label{fig:pE}}
\end{center}
\end{figure}

\begin{figure}
\begin{center}
	\includegraphics[scale=0.6]{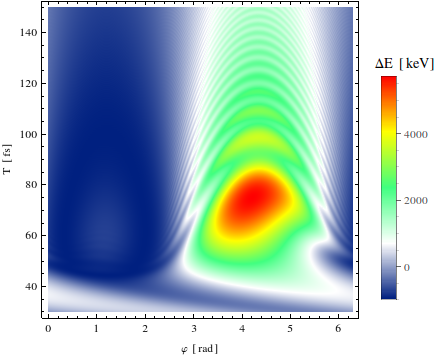}
	\emph{\caption{The dependence of the energy gain on the carrier--envelope phase and the pulse duration. The parameters are $\mathbf{p}_{0} = (-1570 \, \mathrm{keV/c}, 450 \, \mathrm{keV/c}, 0)$, $\lambda = 800 \, \mathrm{nm}$, $I = 10^{17} \, \mathrm{W} \cdot \mathrm{cm}^{-2}$, $\sigma = -0.03886 \, \mathrm{fs}^{-2}$. The optimal values of the carrier--envelope phase and the pulse duration are $\varphi = 4.21 \, \mathrm{rad}$ and $T = 75 \, \mathrm{fs}$, respectively. The oscillations in the positive $T$ direction can be understood as appearing of new optical cycles in the pulse. The local maxima belong to the corresponding sharp rising and falling edges (see also Fig.~\ref{fig:chirped}).}\label{fig:fT}}
\end{center}
\end{figure}

The following three figures show the optimisation of the initial momentum and the plane wave parameters. Their initial values were $\mathbf{p}_{0} = (0, 0, 0)$, $\lambda = 800 \, \mathrm{nm}$, $T = 35 \, \mathrm{fs}$, $\sigma = -0.03886 \, \mathrm{fs}^{-2}$ and $I = 10^{17} \, \mathrm{W} \cdot \mathrm{cm}^{-2}$. \emph{Fig.~\ref{fig:chirped}} shows the electric field which is responsible for the optimal acceleration. \emph{Fig.~\ref{fig:pE}} shows that for  maximal acceleration the electron must not propagate in parallel with the plane wave $\Delta E$ has got a maximum of approximately $550 \, \mathrm{keV}$ by $\mathbf{p}_{0} = (-1570 \, \mathrm{keV/c}, 450 \, \mathrm{keV/c}, 0)$. That is, the ideal injection angle equals to $\alpha = 164^{\circ}$, related to the positive $x$ direction. Applying this initial momentum on the electron we sought the optimal values for the carrier--envelope phase and the pulse duration. The optimal values for these parameters are $\varphi = 4.21 \, \mathrm{rad}$ and $T = 75 \, \mathrm{fs}$, respectively. This further optimisation resulted in a growth by an order of magnitude in the energy gain. The maximal gain by these parameters is $\Delta E = 5\,600 \, \mathrm{keV}$. This is shown on \emph{Fig.~\ref{fig:fT}.} Applying these optimal parameters we optimised the chirp parameter and the intensity of the plane wave. The maximum of the energy gain grew again by an order of magnitude: its maximal value is approximately $\Delta E = 58 \, \mathrm{MeV}$ by $\sigma = -0.03698 \mathrm{fs}^{-2}$ and $I = 10^{21} \mathrm{W} \cdot \mathrm{cm}^{-2}$ and if the other parameters take their optimal values as given above. As a short summary we can say that by properly chosen parameters a single electron is able to gain as much as $58 \, \mathrm{MeV}$ energy from a single plane wave pulse.

\begin{figure}
\begin{center}
	\includegraphics[scale=0.6]{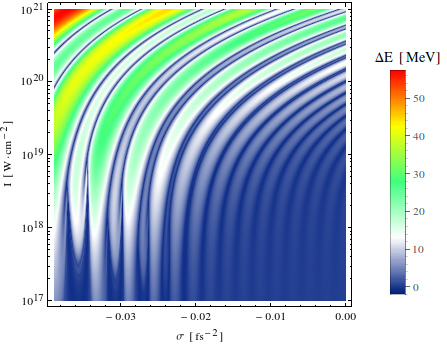}
	\emph{\caption{The dependence of the energy gain on the chirp parameter and the laser intensity. The parameters are $p_{0} = (-1570 \, \mathrm{keV}, 450 \, \mathrm{keV}, 0)$, $\lambda = 800 \, \mathrm{nm}$, $T = 75 \, \mathrm{fs}$, $\varphi = 4.21$. The optimal values of the chirp parameter and the laser intensity are $\sigma = -0.03698 \mathrm{fs}^{-2}$ and $I = 10^{21} \mathrm{W} \cdot \mathrm{cm}^{-2}$, respectively. The oscillations parallel to the horizontal axis arise due the decrement of the chirp parameter. The larger the absolute value of the chirp parameter, the less the number of optical cycles in a single pulse. The local maxima belong to the corresponding sharp rising and falling edges (see also Fig.~\ref{fig:chirped}).}\label{fig:sI}}
\end{center}
\end{figure}

We also investigated the effective theory of the electron acceleration in underdense plasmas. As mentioned earlier, we needed to perform the same analysis by $n_{m} < 1$. We took $n_{p} = 10^{15} \, \mathrm{cm}^{-3}$ which is a typical value for the plasma density in the CERN AWAKE experiment as indicated by \cite{Caldwell}. The index of refraction by this density is $n_{m} = 0.9999997$ which means that the $n_{m} < 1$ case can be quite well approximated with the $n_{m} = 1$ case. Our calculations showed that the energy gain values for these two cases differ less than a per mill from each other.

Finally, we analysed the interaction of a single electron with a short, chirped Gaussian laser pulse. This latter model provides a more realistic description of the acceleration process. \cite{Lax} and \cite{Davis} proved that Gaussian pulses are solutions of Maxwell's equations. \cite{Wang1} also provided a proof and investigated the dynamics of the acceleration process. From a practical point of view, it is known that the state-of-the-art laser systems emit a series of Gaussian laser pulses.

In the effect of the chirp parameter there is a small difference from the case of an electromagnetic plane wave: by a Gaussian pulse we also experienced some gain by zero chirp. However, this gain was negligibly small. This statement is also valid for positive chirp values, hence here we also took only negative values for $\sigma$. There is a major difference in the role of the initial position: since a Gaussian pulse has also got a spatial envelope that causes the laser intensity to decrease rapidly radially we had to place the electron on-axis, far enough for the electron to not to feel the electric field of the laser pulse. We also had to set the initial momentum parallel with the direction of the propagation of the laser pulse so that the electron is able to interact with the pulse as for much time as possible. This means that the initial momentum is parametrized as $\mathbf{p}_{0} = p_{0}\mathbf{e}_{z}$ with $p_{0} \equiv \abs{\mathbf{p}_{0}}$.

We analysed the gain as a function of the initial momentum, the beam waist and the pulse duration. Initially we took the following parameters: $\lambda = 800 \, \mathrm{nm}$, $T = 35 \, \mathrm{fs}$, $I = 10^{21} \, \mathrm{W} \cdot \mathrm{cm}^{-2}$ and $W_{0}$ as some integer multiple of $\lambda$. These are typical values for $\mathrm{Ti}$:Sapphire lasers. Mostly we chose $W_{0} := 100 \mathrm{\lambda}$ because larger focused spot size means less spatial---and temporal---beam divergence.

\begin{figure}
\begin{center}
	\includegraphics[scale=0.6]{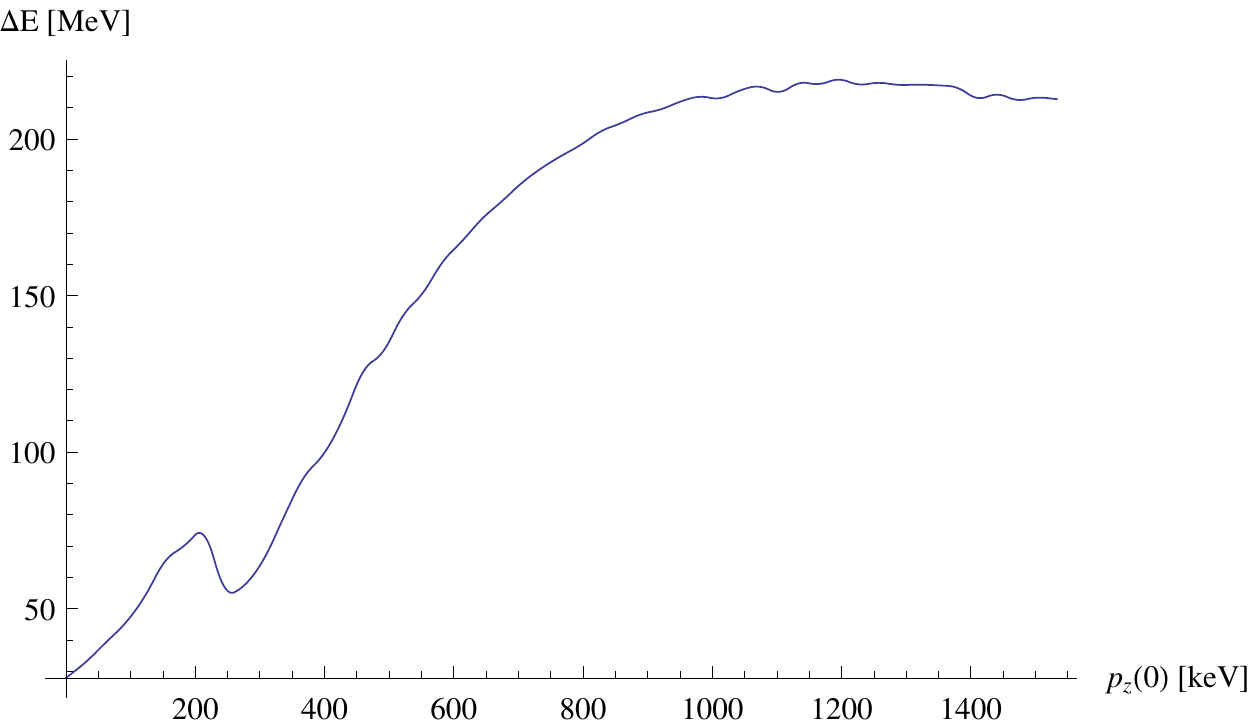}
	\emph{\caption{The energy gain as a function of the initial momentum. $T = 35 \, \mathrm{fs}$, $W_{0} = 100 \lambda$, $I = 10^{21} \, \mathrm{W} \cdot \mathrm{cm}^{-2}$, $\sigma = -0.0194 \, \mathrm{fs}^{-2}$.}\label{fig:pgain-02}}
\end{center}
\end{figure}

\begin{figure}
\begin{center}
	\includegraphics[scale=0.6]{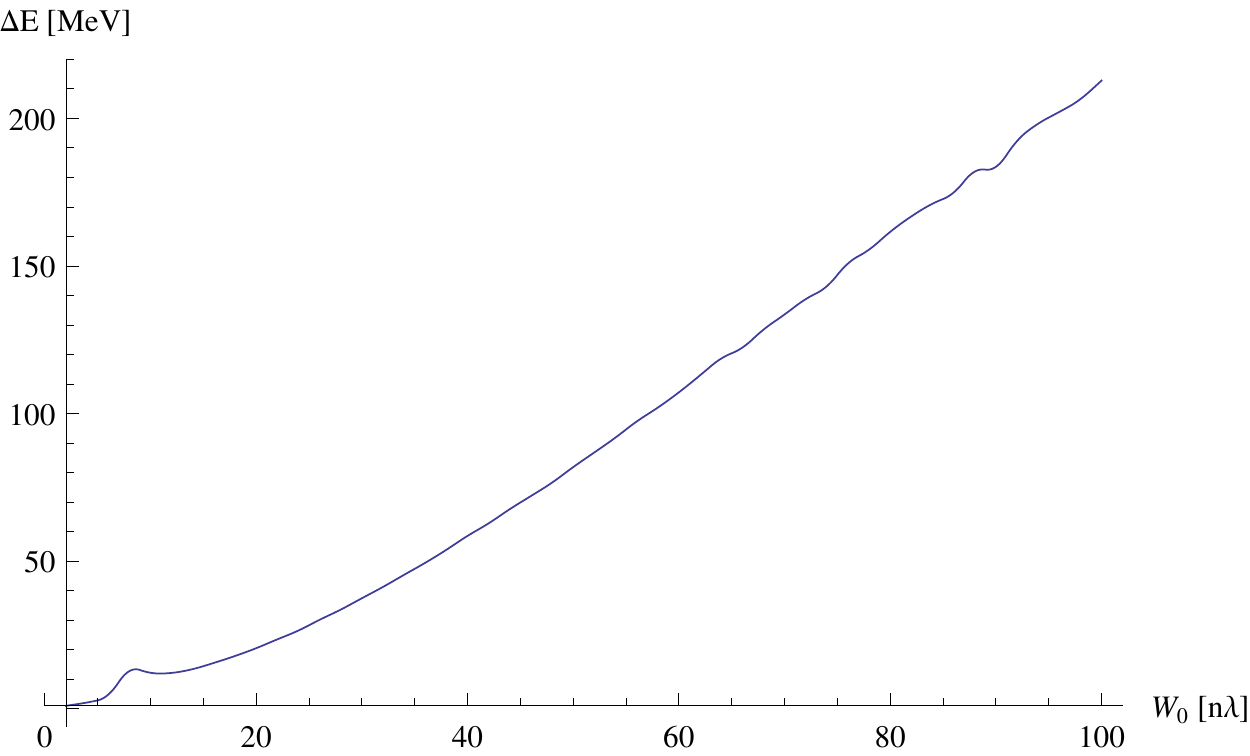}
	\emph{\caption{The energy gain as a function of the beam waist. $p_{0} = 1533 \, \mathrm{keV/c}$, $T = 35 \, \mathrm{fs}$, $I = 10^{21} \, \mathrm{W} \cdot \mathrm{cm}^{-2}$, $\sigma = -0.0194 \, \mathrm{fs}^{-2}$.}\label{fig:wgain-02}}
\end{center}
\end{figure}

\begin{figure}
\begin{center}
	\includegraphics[scale=0.6]{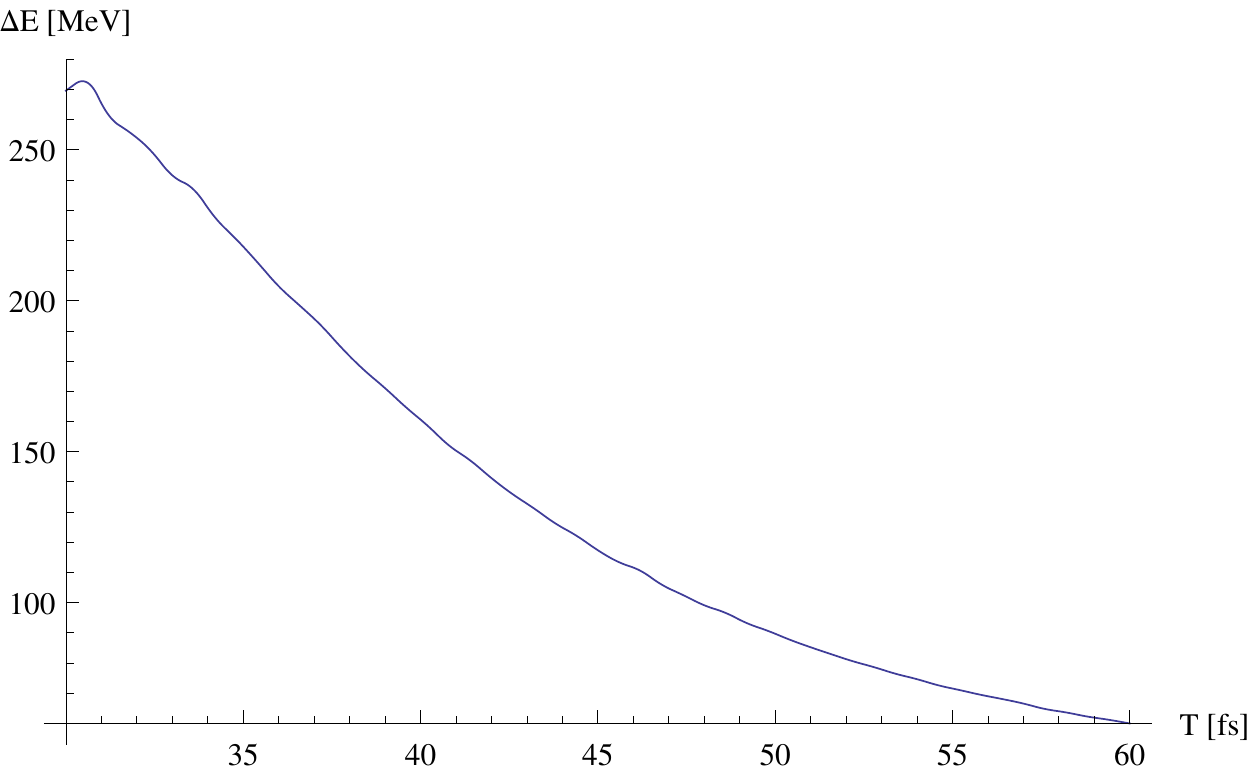}
	\emph{\caption{The energy gain as a function of the pulse duration. $p_{0} = 1533 \, \mathrm{keV/c}$, $W_{0} = 100 \lambda$, $I = 10^{21} \, \mathrm{W} \cdot \mathrm{cm}^{-2}$, $\sigma = -0.0194 \, \mathrm{fs}^{-2}$.}\label{fig:Tgain-02}}
\end{center}
\end{figure}

First we investigated the energy gain as a function of the initial momentum. Our calculations showed that by sufficiently wide focused spot size an electron is able to gain as much as $200 \, \mathrm{MeV}$ energy from a simple pulse. This is shown on \emph{Figs.~\ref{fig:pgain-02} and \ref{fig:wgain-02}.} The energy gain can be increased by shortening the pulse. The dependence of the gain on the pulse duration is presented on \emph{Fig.~\ref{fig:Tgain-02}.} The maximal value of the energy gain from a single pulse by the optimal parameters $p_{0} = 1533 \, \mathrm{keV/c}$, $W_{0} = 100 \lambda$, $T = 31 \, \mathrm{fs}$ and $I = 10^{21} \, \mathrm{W} / \mathrm{cm}^{2}$ is approximately $\Delta E = 275 \, \mathrm{MeV}$. The result presented in this paper agree quite well with other theoretical calculations. Without completeness, see \citep{Sohbatzadeh1, Sohbatzadeh2}.

\newpage
\section{Summary}
In the present paper we studied the electron acceleration with electromagnetic plane waves and Gaussian laser pulses. An effective theory has also been presented for describing the electron acceleration in underdense plasmas. The key idea was the introduction of an effective refraction index for the plasma which can be incorporated into the phase of the laser pulse. Our results have been obtained by numerically solving the relativistic equations of motion. Our calculations showed that in this simple model the acceleration in plasmas can be quite well approximated with the vacuum model, that is, with the $n_{m} = 1$ case.

We performed calculations for the following laser parameters: $\lambda = 800 \, \mathrm{nm}$ wavelength, $T = 30 \, \mathrm{fs}$ pulse duration, $I = 10^{21} \, \mathrm{W} \cdot \mathrm{cm}^{-2}$ intensity and $W_{0} = 100 \lambda$ beam waist (see: \emph{Fig.~\ref{fig:Tgain-02}}). These parameters correspond to an average power of $320 \, \mathrm{TW}$ and a total pulse energy of $E_{\textrm{tot}} = 9.6 \, \mathrm{J}$. The initial injection energy of the electron was $1.5 \, \mathrm{MeV}$ and gained approximately $275 \, \mathrm{MeV}$ energy from a single pulse on an $L_{\textrm{IA}} \approx 5 \, \mathrm{mm}$ interaction length. This determines the acceleration gradient which is $E_{\textrm{acc}} = 58 \, \mathrm{GV/m}$.

We can easily compare these results with the PIC simulations and experimental data obtained by \cite{kneip-near-gev-2009}. They applied a driving laser with $\lambda = 800 \, \mathrm{nm}$ wavelength, $T = 55 \, \mathrm{fs}$ pulse duration, $I = 10^{19} \, \mathrm{W} \cdot \mathrm{cm}^{-2}$ intensity and $W_{0} = 10 \mathrm{\mu m}$ beam waist on a plasma with plasma density of $n_{e} = 5.5 \cdot 10^{18} \, \mathrm{cm}^{-3}$. The parameters mentioned above correspond to an average power of $P = 180 \, \mathrm{TW}$ and total pulse energy of $E_{\textrm{tot}} = 10 \, \mathrm{J}$. For the interaction lengths of $L_{\textrm{IA}} = 5 \, \mathrm{mm}$ and $L_{\textrm{IA}} = 10 \, \mathrm{mm}$, the final energies of the electrons were $E_{\textrm{final}} = 420 \, \mathrm{MeV}$ and $E_{\textrm{final}} = 800 \, \mathrm{MeV}$, respectively. The calculated accelerating gradient was $E_{\textrm{acc}} = 80 \, \mathrm{GV/m}$.

Of course, our single-particle effective model cannot take into account precisely the non-linear plasma effects and additional collective phenomena. Still, it is worthy to mention that this simple phenomenological model compares quite reasonably with the sophisticated PIC simulations and experimental data. Our results agree with that obtained by \cite{kneip-near-gev-2009} within a factor of 2 both for the accelerating gradient and the maximal energy gain.

\section*{Acknowledgement}
S.~V.~has been supported by the National Scientific Research Foundation OTKA, Grant No.~K 104260. Partial support by the ELI-ALPS project is also acknowledged. The ELI-ALPS project (GOP-1.1.1-12/B-2012-0001) is supported by the European Union and co-financed by the European Regional Development Fund.

\bibliographystyle{lpb}
\bibliography{references}{}

\begin{thebibliography}{27}
\providecommand{\natexlab}[1]{#1}
\providecommand{\url}[1]{\texttt{#1}}
\providecommand{\urlprefix}{URL }
\providecommand{\selectlanguage}[1]{\relax}
\providecommand{\eprint}[2][]{\url{#2}}

\bibitem[{Abramowitz and Stegun(1972)Abramowitz \& Stegun}]{Abramowitz}
\textsc{Abramowitz, M.} and \textsc{Stegun, I.A.} (1972).
\newblock \textit{Handbook of mathematical functions}, \textit{Applied
  Mathematics Series}, vol.~55, chap.~7.
\newblock Washington, D.C. 20402: U.S. Government Printing Office, 10 ed.
\newblock (Equations (7.3.1)--(7.3.4)).

\bibitem[{Cheng and Xu(1999)Cheng \& Xu}]{Cheng}
\textsc{Cheng, Y.} and \textsc{Xu, Z.} (1999).
\newblock Vacuum laser acceleration by an ultrashort, high-intensity laser
  pulse with a sharp rising edge.
\newblock \textit{Appl. Phys. Lett.} \textbf{74}, 2116--2118.

\bibitem[{Davis(1979)}]{Davis}
\textsc{Davis, L.W.} (1979).
\newblock Theory of electromagnetic beams.
\newblock \textit{Phys. Rev. A} \textbf{19}, 1177--1179.

\bibitem[{Esarey \textit{et~al.}(2009)Esarey, Schroeder \& Leemans}]{Esarey2}
\textsc{Esarey, E.}, \textsc{Schroeder, C.B.} and \textsc{Leemans, W.P.}
  (2009).
\newblock Physics of {L}aser-{D}riven {P}lasma-{B}ased {E}lectron
  {A}ccelerators.
\newblock \textit{Rev.~Mod.~Phys.} \textbf{81}, 1229--1285.

\bibitem[{Fonseca \textit{et~al.}(2002)Fonseca, Silva, Tsung, Decyk, Lu, Ren,
  Mori, Deng, Lee, Katsouleas \& Adam}]{OSIRIS}
\textsc{Fonseca, R.A.}, \textsc{Silva, L.O.}, \textsc{Tsung, F.S.},
  \textsc{Decyk, V.K.}, \textsc{Lu, W.}, \textsc{Ren, C.}, \textsc{Mori, W.B.},
  \textsc{Deng, S.}, \textsc{Lee, S.}, \textsc{Katsouleas, T.} and
  \textsc{Adam, J.C.} (2002).
\newblock {\selectlanguage{english}OSIRIS: A {T}hree-{D}imensional, {F}ully
  {R}elativistic {P}article in {C}ell {C}ode for {M}odeling {P}lasma {B}ased
  {A}ccelerators}.
\newblock In \textit{Computational {S}cience -- {ICCS} 2002}, \textit{Lecture
  {N}otes in {C}omputer {S}cience}, vol. 2331 (P.M.A. Sloot, A.G. Hoekstra,
  C.J.K. Tan \& J.J. Dongarra, Eds.), pp. 342--351. Springer Berlin Heidelberg.

\bibitem[{Geddes \textit{et~al.}(2005)Geddes, Toth, van Tilborg, Esarey,
  Schroeder, Cary \& Leemans}]{Geddes}
\textsc{Geddes, C.G.R.}, \textsc{Toth, C.}, \textsc{van Tilborg, J.},
  \textsc{Esarey, E.}, \textsc{Schroeder, C.B.}, \textsc{Cary, J.} and
  \textsc{Leemans, W.P.} (2005).
\newblock Guiding of {R}elativistic {L}aser {P}ulses by {P}reformed {P}lasma
  {C}hannels.
\newblock \textit{Phys.~Rev.~Lett.} \textbf{95}, 145002 (pages~4).

\bibitem[{Gonsalves \textit{et~al.}(2011)Gonsalves, Nakamura, Lin, Panasenko,
  Shiraishi, Sokollik, Benedetti, Schroeder, Geddes, van Tilborg, Osterhoff,
  Esarey, Toth \& Leemans}]{Gonsalves}
\textsc{Gonsalves, A.J.}, \textsc{Nakamura, K.}, \textsc{Lin, C.},
  \textsc{Panasenko, D.}, \textsc{Shiraishi, S.}, \textsc{Sokollik, T.},
  \textsc{Benedetti, C.}, \textsc{Schroeder, C.}, \textsc{Geddes, C.G.R.},
  \textsc{van Tilborg, J.}, \textsc{Osterhoff, J.}, \textsc{Esarey, E.},
  \textsc{Toth, C.} and \textsc{Leemans, W.P.} (2011).
\newblock Tunable laser plasma accelerator based on longitudinal density
  tailoring.
\newblock \textit{Nat.~Phys} \textbf{7}, 862--866.

\bibitem[{Kneip \textit{et~al.}(2009)Kneip, Nagel, Martins, Mangles, Bellei,
  Chekhlov, Clarke, Delerue, Divall, Doucas, Ertel, Fiuza, Fonseca, Foster,
  Hawkes, Hooker, Krushelnick, Mori, Palmer, Phuoc, Rajeev, Schreiber,
  Streeter, Urner, Vieira, Silva \& Najmudin}]{kneip-near-gev-2009}
\textsc{Kneip, S.}, \textsc{Nagel, S.R.}, \textsc{Martins, S.F.},
  \textsc{Mangles, S.P.D.}, \textsc{Bellei, C.}, \textsc{Chekhlov, O.},
  \textsc{Clarke, R.J.}, \textsc{Delerue, N.}, \textsc{Divall, E.J.},
  \textsc{Doucas, G.}, \textsc{Ertel, K.}, \textsc{Fiuza, F.}, \textsc{Fonseca,
  R.}, \textsc{Foster, P.}, \textsc{Hawkes, S.J.}, \textsc{Hooker, C.J.},
  \textsc{Krushelnick, K.}, \textsc{Mori, W.B.}, \textsc{Palmer, C.A.J.},
  \textsc{Phuoc, K.T.}, \textsc{Rajeev, P.P.}, \textsc{Schreiber, J.},
  \textsc{Streeter, M.J.V.}, \textsc{Urner, D.}, \textsc{Vieira, J.},
  \textsc{Silva, L.O.} and \textsc{Najmudin, Z.} (2009).
\newblock Near-{GeV} Acceleration of Electrons by a Nonlinear Plasma Wave
  Driven by a Self-Guided Laser Pulse.
\newblock \textit{Phys. Rev. Lett.} \textbf{103}, 035\,002.

\bibitem[{Lax \textit{et~al.}(1975)Lax, Louisell \& McKnight}]{Lax}
\textsc{Lax, M.}, \textsc{Louisell, W.H.} and \textsc{McKnight, W.B.} (1975).
\newblock From Maxwell to paraxial wave optics.
\newblock \textit{Phys. Rev. A} \textbf{11}, 1365--1370.

\bibitem[{Lifschitz \textit{et~al.}(2006)Lifschitz, Faure, Glinec, Malka \&
  Mora}]{lifschitz-proposed-2006}
\textsc{Lifschitz, A.}, \textsc{Faure, J.}, \textsc{Glinec, Y.}, \textsc{Malka,
  V.} and \textsc{Mora, P.} (2006).
\newblock Proposed scheme for compact {GeV} laser plasma accelerator.
\newblock \textit{Laser and Particle Beams} \textbf{24}, 255--259.

\bibitem[{Malka \textit{et~al.}(2002)Malka, Fritzler, Lefebvre, Aleonard,
  Burgy, Chambaret, Chemin, Krushelnick, Malka, Mangles, Najmudin, Pittman,
  Rousseau, Scheurer, Walton \& Dangor}]{Malka}
\textsc{Malka, V.}, \textsc{Fritzler, S.}, \textsc{Lefebvre, E.},
  \textsc{Aleonard, M.M.}, \textsc{Burgy, F.}, \textsc{Chambaret, J.P.},
  \textsc{Chemin, J.F.}, \textsc{Krushelnick, K.}, \textsc{Malka, G.},
  \textsc{Mangles, S.P.D.}, \textsc{Najmudin, Z.}, \textsc{Pittman, M.},
  \textsc{Rousseau, J.P.}, \textsc{Scheurer, J.N.}, \textsc{Walton, B.} and
  \textsc{Dangor, A.E.} (2002).
\newblock Electron {A}cceleration by a {W}ake {F}ield {F}orced by an {I}ntense
  {U}ltrashort {L}aser {P}ulse.
\newblock \textit{Science} \textbf{298}, 1596--1600.

\bibitem[{Nakajima \textit{et~al.}(1995)Nakajima, Fisher, Kawakubo, Nakanishi,
  Ogata, Kato, Kitagawa, Kodama, Mima, Shiraga, Suzuki, Yamakawa, Zhang,
  Sakawa, Shoji, Nishida, Yugami, Downer \& Tajima}]{Nakajima}
\textsc{Nakajima, K.}, \textsc{Fisher, D.}, \textsc{Kawakubo, T.},
  \textsc{Nakanishi, H.}, \textsc{Ogata, A.}, \textsc{Kato, Y.},
  \textsc{Kitagawa, Y.}, \textsc{Kodama, R.}, \textsc{Mima, K.},
  \textsc{Shiraga, H.}, \textsc{Suzuki, K.}, \textsc{Yamakawa, K.},
  \textsc{Zhang, T.}, \textsc{Sakawa, Y.}, \textsc{Shoji, T.}, \textsc{Nishida,
  Y.}, \textsc{Yugami, N.}, \textsc{Downer, M.} and \textsc{Tajima, T.} (1995).
\newblock Observation of {U}ltrahigh {G}radient {E}lectron {A}cceleration by a
  {S}elf-{M}odulated {I}ntense {S}hort {L}aser {P}ulse.
\newblock \textit{Phys.~Rev.~Lett.} \textbf{74}, 4428--4431.

\bibitem[{Pocsai(2014)}]{Pocsai}
\textsc{Pocsai, M.A.} (2014).
\newblock \textit{R\'{e}szecskegyors\'{\i}t\'{a}s l\'{e}zerrel}.
\newblock Master's thesis, Roland E\"{o}tv\"{o}s University.
\newblock (Unpublished).

\bibitem[{Pukhov and Meyer-ter Vehn(2002)Pukhov \& Meyer-ter Vehn}]{Pukhov}
\textsc{Pukhov, A.} and \textsc{Meyer-ter Vehn, J.} (2002).
\newblock {\selectlanguage{english}Laser {W}ake {F}ield {A}cceleration: {T}he
  {H}ighly {N}on-{L}inear {B}roken-{W}ave {R}egime}.
\newblock \textit{Appl.~Phys.~B} \textbf{74}, 355--361.

\bibitem[{Rosenzweig \textit{et~al.}(1991)Rosenzweig, Breizman, Katsouleas \&
  Su}]{rosenzweig-acceleration-1991}
\textsc{Rosenzweig, J.B.}, \textsc{Breizman, B.}, \textsc{Katsouleas, T.} and
  \textsc{Su, J.J.} (1991).
\newblock Acceleration and focusing of electrons in two-dimensional nonlinear
  plasma wake fields.
\newblock \textit{Phys. Rev. A} \textbf{44}, R6189--R6192.

\bibitem[{Sohbatzadeh and Aku(2011)Sohbatzadeh \& Aku}]{Sohbatzadeh2}
\textsc{Sohbatzadeh, F.} and \textsc{Aku, H.} (2011).
\newblock Polarization {E}ffect of a {C}hirped {F}aussian {L}aser {P}ulse on
  the {E}lectron {B}unch {A}cceleration.
\newblock \textit{Journal of Plasma Physics} \textbf{77}, 39--50.

\bibitem[{Sohbatzadeh \textit{et~al.}(2006)Sohbatzadeh, Mirzanejhad \&
  Ghasemi}]{Sohbatzadeh1}
\textsc{Sohbatzadeh, F.}, \textsc{Mirzanejhad, S.} and \textsc{Ghasemi, M.}
  (2006).
\newblock Electron {A}cceleration by a {C}hirped {G}aussian {L}aser {P}ulse in
  {V}acuum.
\newblock \textit{Phys.~Plasmas} \textbf{13}, 123108.

\bibitem[{Strickland and Mourou(1985)Strickland \& Mourou}]{cpa2}
\textsc{Strickland, D.} and \textsc{Mourou, G.} (1985).
\newblock Compression of {A}mplified {C}hirped {O}ptical {P}ulses.
\newblock \textit{Opt.~Commun.} \textbf{56}, 219--221.

\bibitem[{Tajima and Dawson(1979)Tajima \& Dawson}]{Tajima}
\textsc{Tajima, T.} and \textsc{Dawson, J.M.} (1979).
\newblock Laser {E}lectron {A}ccelerator.
\newblock \textit{Phys.~Rev.~Lett.} \textbf{43}, 267--270.

\bibitem[{Varr\'{o}(2007)}]{varro-linear-2007}
\textsc{Varr\'{o}, S.} (2007).
\newblock Linear and nonlinear absolute phase effects in interactions of
  ulrashort laser pulses with a metal nano-layer or with a thin plasma layer.
\newblock \textit{Laser and Particle Beams} \textbf{25}, 379--390.

\bibitem[{Varr{\'{o}}(2014)}]{VarroNm}
\textsc{Varr{\'{o}}, S.} (2014).
\newblock New {E}xact {S}olutions of the {D}irac and {K}lein--{G}ordon
  {E}quations of a {C}harged {P}article {P}ropagating in a {S}trong {L}aser
  {F}ield in an {U}nderdense {P}lasma.
\newblock \textit{Nuc.~Instr.~Meth.~Phys.~Res.~A} \textbf{740}, 280--283.
\newblock Proceedings of the {F}irst {E}uropean {A}dvanced {A}ccelerator
  {C}oncepts {W}orkshop 2013.

\bibitem[{Varr\'{o} and Farkas(2008)Varr\'{o} \&
  Farkas}]{varro-attosecond-2008}
\textsc{Varr\'{o}, S.} and \textsc{Farkas, G.} (2008).
\newblock Attosecond electron pulses from interference of above-threshold de
  {Broglie} waves.
\newblock \textit{Laser and Particle Beams} \textbf{26}, 9--20.

\bibitem[{Varr{\'{o}} and Kocsis(1992)Varr{\'{o}} \& Kocsis}]{VarroExact}
\textsc{Varr{\'{o}}, S.} and \textsc{Kocsis, G.} (1992).
\newblock Classical {M}otion of a {C}harged {P}article in the {P}resence of a
  {S}tatic, {H}omogenous {M}agnetic {F}ield and a {L}inearly {F}requency
  {S}hifted {E}lectromagnetic {P}lanewave.
\newblock In \textit{Proceedings of the {T}hird {E}uropean {P}article
  {A}ccelerator {C}onference}, \textit{EPAC 92}, vol.~2 (H.~Henke, H.~Homeyer,
  C.~Petit-Jean-Genaz \& G.~sur Yvette, Eds.), pp. 964--966. Editions
  Fronti{\`e}res.

\bibitem[{Vieira and Mendon{\c{c}}a(2014)Vieira \& Mendon{\c{c}}a}]{Vieira}
\textsc{Vieira, J.} and \textsc{Mendon{\c{c}}a, J.T.} (2014).
\newblock Nonlinear {L}aser {D}riven {D}onut {W}akefields for {P}ositron and
  {E}lectron {A}cceleration.
\newblock \textit{Phys.~Rev.~Lett.} \textbf{112}, 215001 (pages~5).

\bibitem[{Wang \textit{et~al.}(1999)Wang, Ho, Feng, Kong, Wang, Yuan \&
  Scheid}]{Wang1}
\textsc{Wang, J.X.}, \textsc{Ho, Y.K.}, \textsc{Feng, L.}, \textsc{Kong, Q.},
  \textsc{Wang, P.X.}, \textsc{Yuan, Z.S.} and \textsc{Scheid, W.} (1999).
\newblock High-intensity laser-induced electron acceleration in vacuum.
\newblock \textit{Phys. Rev. E} \textbf{60}, 7473--7478.

\bibitem[{Wang \textit{et~al.}(2000)Wang, Scheid, Hoelss \& Ho}]{Wang2}
\textsc{Wang, J.}, \textsc{Scheid, W.}, \textsc{Hoelss, M.} and \textsc{Ho, Y.}
  (2000).
\newblock Electron acceleration by intense shock-like laser pulses in vacuum.
\newblock \textit{Physics Letters A} \textbf{275}, 323--328.

\bibitem[{Xia \textit{et~al.}(2011)Xia, Caldwell, Huang \& Mori}]{Caldwell}
\textsc{Xia, G.}, \textsc{Caldwell, A.}, \textsc{Huang, C.} and \textsc{Mori,
  W.B.} (2011).
\newblock Simulation {S}tudy on {P}roton-{D}riven {PWFA} {B}ased on {CERN}
  {SPS} {B}eam.
\newblock In \textit{Proceedings of 2011 {P}article {A}ccelerator {C}onference,
  {N}ew {Y}ork, {NY}, {USA}}, pp. 301--303.

\end{thebibliography}
\end{document}